\documentclass[aps,prl,showpacs,notitlepage,twocolumn,superscriptaddress,nofootinbib,preprintnumbers]{revtex4-2}

\usepackage{graphicx}
\usepackage{bm}
\usepackage{hyperref}
\usepackage{amsmath} 
\usepackage{amssymb}
\usepackage[table]{xcolor}
\usepackage{array}
\usepackage{multirow}
\usepackage{physics}
\usepackage{lipsum}
\usepackage{float}

\newcommand{\ua}{\uparrow}
\newcommand{\da}{\downarrow}
\newcommand{\br}{\bold{r}}
\newcommand{\psit}{\psi_{\theta}}
\newcommand{\mo}{moir\'e }

\begin{document}
\title{Artificial intelligence for artificial materials: moir\'e atom}
\author{Di Luo}
\affiliation{The NSF AI Institute for Artificial Intelligence and Fundamental Interactions}
\affiliation{Center for Theoretical Physics, Massachusetts Institute of Technology, Cambridge, MA 02139}
\affiliation{Department of Physics, Harvard University, Cambridge, MA 02138, USA}
\author{Aidan P. Reddy}
\author{Trithep Devakul}
\author{Liang Fu}
\affiliation{Department of Physics, Massachusetts Institute of Technology, Cambridge, Massachusetts 02139, USA}

\begin{abstract}
Moiré engineering in atomically thin van der Waals heterostructures creates   artificial quantum materials with designer properties.  
We solve the many-body problem of interacting electrons confined to a  
moiré superlattice potential minimum (the moir\'e atom) using a 2D fermionic neural network. We show that strong Coulomb interactions in combination with the anisotropic moiré potential lead to striking ``Wigner molecule" charge density distributions observable with scanning tunneling microscopy.
\end{abstract}

\maketitle

The study of atoms has been a cornerstone of physics, chemistry and material science for centuries. Quantum theory of electrons in atoms provides the foundation for electronic structure theory of solids. 
In the last few decades, advances in semiconductor technology have enabled the fabrication of  artificial atoms, i.e., quantum dots containing a tunable number of electrons. In recent years, the advent of two-dimensional (2D) semiconductors and moir\'e heterostructures offers an exciting new platform to create artificial solids, periodic arrays of ``moir\'e atoms'' with a period  much larger than the angstrom scale. 
By tuning the density of mobile electrons with electrostatic gating, a wide variety of quantum phases of matter has been discovered, including Mott insulators \cite{regan2020mott, tang2020simulation}, electron Wigner crystals \cite{li2021imaging}, the quantum anomalous Hall state \cite{li2021quantum}, and excitonic insulators \cite{gu2022dipolar, zhang2022correlated}.  

Direct first-principles study of moir\'e quantum materials is difficult due to a multitude of energy/length scales and many-body correlation effects involved. Since the moir\'e period is much larger than the atomic spacing, the low-energy physics is captured by effective continuum Hamiltonians that involve coarse-grained degrees of freedom only. 
The simplest continuum Hamiltonian, which describes various semiconductor heterobilayers and twisted homobilayers, is that of a 2D electron system with Coulomb interaction and a slowly-varying periodic potential with the periodicity of the moiré lattice \cite{wu2018hubbard, zhang2020moire, zhang2020density}. This minimal model for semiconductor moir\'e materials is a generalization of the uniform electron gas (jellium model), and at the same time, a simplification of real solids in which the crystal potential is singular near the nucleus. Despite its simple form, we believe that this model of interacting electrons encompasses a myriad of phases that remain to be explored.

Recent advance in artificial intelligence has opened up new opportunities for quantum many-body physics. The insight comes from designing efficient neural network representation for quantum many-body wave function, which is first proposed as the neural network quantum state in Ref.~\cite{Carleo602}. It has been shown that neural network quantum states are able to represent a large family of wave functions including highly entangled states and have higher expressivity than tensor networks~\cite{Deng_2017, gao2017efficient, Glasser_2018, Levine_2019, sharir2021neural, luo_inf_nnqs}. Promising applications include solving ground state properties~\cite{robledo2022fermionic, chen2022simulating, doi:10.1126/science.aag2302, Hibat_Allah_2020, PhysRevLett.124.020503, Irikura_2020, PhysRevResearch.3.023095, Han_2020,ferminet,Choo_2019,rnn_wavefunction,paulinet,Glasser_2018,Stokes_2020,Nomura_2017,martyn2022variational,Luo_2019,PhysRevLett.127.276402, https://doi.org/10.48550/arxiv.2101.07243,luo2022gauge} and simulating finite temperature and real time dynamics \cite{xie2021ab,wang2021spacetime,py2021, gutierrez2020real, Schmitt_2020,Vicentini_2019,PhysRevB.99.214306,PhysRevLett.122.250502,PhysRevLett.122.250501,luo_gauge_inv,luo_povm}. In particular, fermionic neural network wave functions have demonstrated great potential in condensed matter and quantum chemistry. The development originates from the neural network backflow~\cite{Luo_2019}, which utilizes neural network parameterization of many-body determinants 
to capture the anti-symmetry property of many-electron wavefunctions and generalize the Slater-Jastrow-Backflow hierarchy. FermiNet~\cite{ferminet} and PauliNet~\cite{paulinet} have developed the idea into continuous space and later advancement has made great progresses in electronic structure calculations~\cite{cassella2023discovering,wilson2021simulations,li2022fermionic,entwistle2023electronic,scherbela2022solving}.

In this work, we apply a 2D fermionic neural network based on FermiNet to 
solve for the ground states of moir\'e atoms 
and obtain images of electron density. 
Working directly with the continuum model of interacting electrons in a moir\'e atom, 
our calculation predicts striking charge density profiles of multi-electron moiré atoms that 
differ greatly from those of literal atoms. 
Originating from Coulomb interaction and moir\'e crystal anisotropy, these density profiles  
can be directly visualized with scanning tunneling microscopy. 
Our work demonstrates the power of neural-network based variational methods in combination with multi-scale modeling in studying artificial materials. It allows us to solve the many-electron Schr\"{o}dinger equation of an effective low-energy model for semiconductor moir\'e materials, with model parameters determined by \textit{ab initio} density functional theory calculations at charge neutrality \cite{zhang2020moire, angeli2021gamma, zhang2021electronic, zhang2021spin}.  
Our work paves the way for future applications of artificial intelligence in understanding and designing artificial materials.

Electrons (holes) doped into the conduction (valence) band of many semiconductor moiré superlattices are described by an effective continuum Hamiltonian with a periodic potential and a Coulomb interaction~\cite{wu2018hubbard,zhang2020density}. 
\begin{equation}
\begin{split}
    \mathcal{H} &\equiv T+V+U\\
&=\sum_{i}\left(\frac{\bm{p}_i^2}{2m}+V(\bm{r}_i)\right) + \sum_{j\neq i}\frac{e^2}{\epsilon|\bm{r}_i-\bm{r}_j|},
\end{split}\label{eq:contham}
\end{equation}
where $m$ is the effective mass and $V(\bm{r})$ is a smooth potential having the periodicity of the moir\'e superlattice, hereafter referred to as the moir\'e potential.
The minima of $V{(\bm{r})}$ typically form a triangular lattice of ``moir\'e atoms" with the moir\'e period $a_M$.

\begin{figure}[t]
    \centering
\includegraphics[width=\columnwidth]{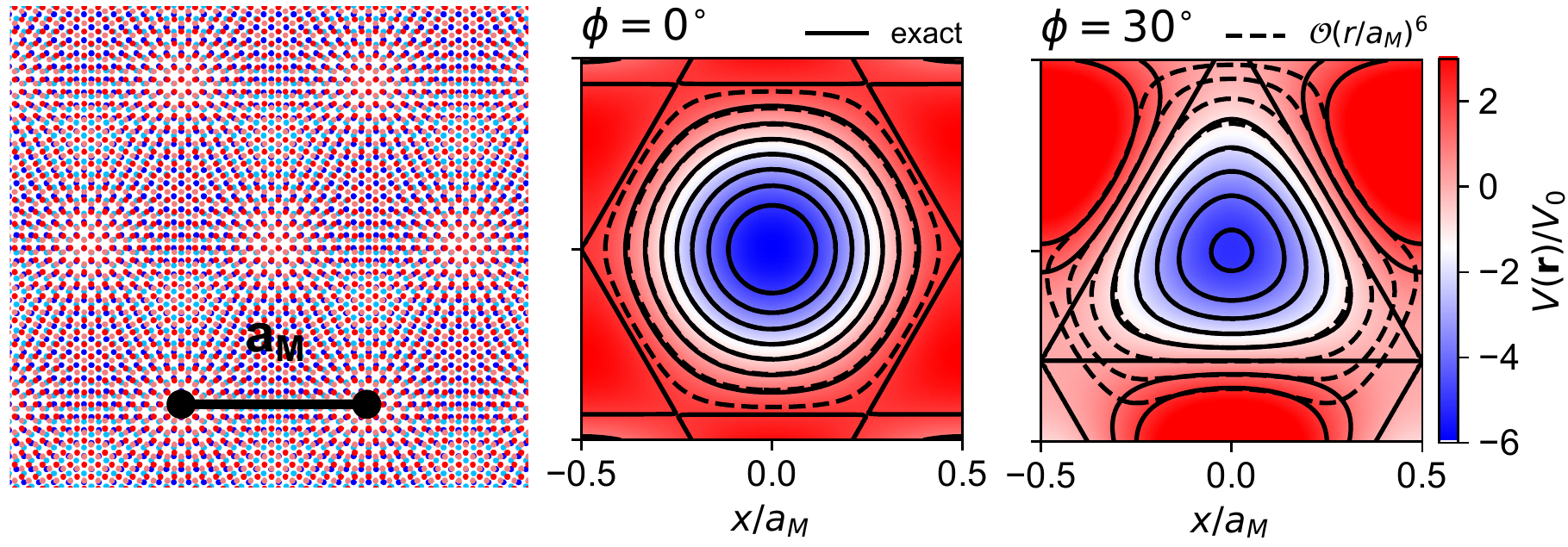}
    \caption{Depiction of semiconductor moiré superlattice (left). Color plots of the moiré potential with phase parameter $\phi=0^{\circ}$ (middle) and $\phi=30^{\circ}$ (right), including exact and approximate  
    equipotentials. Away form the potential minimum, the equipotentials evolve from circular to hexagonal and triangular respectively.} 
\label{fig:moireCartoon}
\end{figure}

Solving this many-body model numerically has proven challenging.
Exact diagonalization study is limited to small system size and truncation to a few bands  \cite{morales2021metal, li2021spontaneous, morales2022nonlocal}. Consequently, it quickly becomes unfeasible at higher electron densities and at realistic interaction strength where band mixing is significant. Hartree-Fock theory neglects quantum correlation effects which are essential to describe highly-entangled states \cite{hu2021competing}. Density functional method with 2D exchange and correlation energy functional has recently been developed to  solve the continuum model \cite{zhang2020density}. While being accurate in the high density regime, it suffers from the self-interaction error in the low density and strong potential regime.       
Much theoretical analysis has proceeded by  downfolding the continuum model to a one (or two) band Hubbard model \cite{wu2018hubbard, zhang2020moire, pan2020quantum, zang2021hartree, devakul2021magic, devakul2022quantum, zhou2022quantum}. This simplified description succeeds in describing several phenomena including incompressible states at fractional fillings known as generalized Wigner crystals. However, as we showed recently, this approach inevitably fails at large moir\'e period where Coulomb interaction necessarily exceeds the band gap \cite{reddy2023moir}.

\emph{Moir\'e atoms ---}
We proceed from the observation that, when moiré atoms are well isolated (flat band limit), the most important electronic interactions in the system are those that occur within a given moiré atom. 
To model an atom in isolation, 
we expand the moir\'e potential about its minimum at the origin, 
\begin{align}\label{eq:atomTaylor}
        V(\bm{r}) \approx \frac{1}{2}k r^2+c_3\sin(3\theta)r^3+\dots
\end{align}
up to a constant term, where the coefficients are determined by the moir\'e potential landscape.
As our original moir\'e potential, we use the first harmonic moir\'e potential~\cite{wu2018hubbard} (see Appendix)
which is characterized by two quantities, an overall potential strength $V_0$ and a phase parameter $\phi$ which controls the potential landscape.
Henceforth, we therefore consider the Hamiltonian for an isolated moir\'e atom given by Eq~\ref{eq:contham} with a moir\'e potential $V(\bm{r})$ truncated to order $\mathcal{O}((r/a_M)^6)$ about the origin.
The result is a circular oscillator with frequency $\omega=\sqrt{k/m}$ and high-order anisotropic corrections that we refer to collectively as the moiré crystal field. 

Two representative moir\'e potentials are shown in Fig~\ref{fig:moireCartoon}, obtained from first harmonic moir\'e potentials 
with phase parameters $\phi=0^\circ$ and $\phi=30^\circ$. The 
potential contour clearly deviates from circular at distance $r\sim a_M$, the only length scale in 
$V(\bm r)$ within the first harmonic approximation. 
The $\phi=0^\circ$ potential has a weaker moir\'e crystal field effect and features hexagonal equipotentials, while the $\phi=30^\circ$ potential features a stronger trigonal crystal field and triangular equipotentials. 
As we will see, the moiré crystal field plays the important role of enabling the moiré atoms to have highly anisotropic charge density profiles when interactions are strong.
The value of $\phi$ is generally material dependent. 
For transition metal dichalcogenide (TMD)
heterobilayers (WSe$_2$/WS$_2$, MoSe$_2$/WSe$_2$, etc), first-principles calculations find 
$\phi$ varies greatly from $\sim 0^{\circ}$ to $45^{\circ}$ \cite{zhang2020moire, kometter2022hofstadter}. For AA stacked $\Gamma$-valley twisted homobilayers, the moir\'e potential is $D_6$ symmetric with $\phi=60^\circ$\cite{angeli2021gamma}.
 We will focus primarily on the potential with $\phi=30^\circ$, $V_0=20$meV, 
at a realistic dielectric constant $\epsilon=10$. 

In Ref~\cite{reddy2023moir}, it was shown that the continuum model Hamiltonian can be characterized by three important length scales: 1) the moiré lattice constant $a_M$, 2) the quantum confinement length $\xi_{0} \equiv \left( \hbar^2/(mk) \right)^{1/4}$, i.e.  
the width of the gaussian profile of the harmonic oscillator wavefunctions, and 3) the Coulomb confinement length $\xi_{c} = \left(\frac{e^2}{4\epsilon k}\right)^{1/3}$, the equilibrium radius of two classical point charges diametrically opposed about the origin of the harmonic oscillator.
The approximation of an isolated atom is justified when $a_M\gg \xi_0,\xi_c$ and is internally consistent as long as the size of the many-body electron ground state is smaller than the inter-atomic spacing.

Within each moir\'e atom, 
the ratio $\frac{e^2/(\epsilon \xi_0)}{\hbar\omega} = 4(\xi_c/\xi_0)^3 \equiv \lambda$ characterizes the relative importance of interaction and single-particle energies.
When $\xi_c \ll \xi_0$, each moiré atom is a weakly-interacting few body system, providing a two-dimensional realization of Schrodinger's model of the atom. The evolution of its electronic structure with particle number can be understood in terms of filling of successive energetic shells. The lifting of degeneracies due to interactions can be understood perturbatively, for instance through Hund's rules \cite{tarucha1996shell}. 

On the other hand, when $\xi_0 \ll \xi_c$, each moiré atom is a strongly-interacting few body system. In combination with the crystal field, Coulomb interactions drive the system to more closely resemble Thomson's plum pudding model of the atom in which classical point charges exist within a neutralizing background. The resultant state is known as a ``Wigner molecule", in analogy to the Wigner crystal phase of the homogenous electron gas \cite{egger1999crossover, filinov2001wigner, kalliakos2008molecular, yannouleas1999spontaneous, ghosal2006correlation}.

We now compare the physical parameters of typical moiré atoms with those of typical GaAs quantum dots. In GaAs electrostatic quantum dots, the effective mass is $m=0.067m_e$, the oscillator frequency $\approx 2-4$ meV, corresponding to a range of quantum confinement lengths $\xi_0 \approx 17-24$nm \cite{maksym2000molecular}. With a dielectric constant of $\epsilon = 12.4$, the corresponding coulomb coupling constants range from $\lambda  \approx 1.7-2.4$. In a typical moiré atom of TMD heterostructures, the effective mass is $\approx 0.5-1.5 m_e$, the oscillator frequency is several tens of meVs, and the quantum confinement length is $\xi_0\approx 1-2\text{nm}$. Assuming a range of dielectric environments with $\epsilon=5-10$, the corresponding Coulomb coupling constant $\lambda \approx 1.5-8$.  Due to the much heavier effective mass and reduced screening in two dimensions,  moiré atoms are clearly distincguished from GaAs quantum dots by their small size, large energy scale and stronger Couloumb coupling. Moreover, the confinement potential in moir\'e atoms becomes significantly non-parabolic and non-isotropic at $r \sim a_M$, an additional length scale that is absent in parabolic quantum dot~\cite{tavernier2004correlation,bedanov1994ordering}.

\begin{figure}
    \centering
\includegraphics[width=\columnwidth]{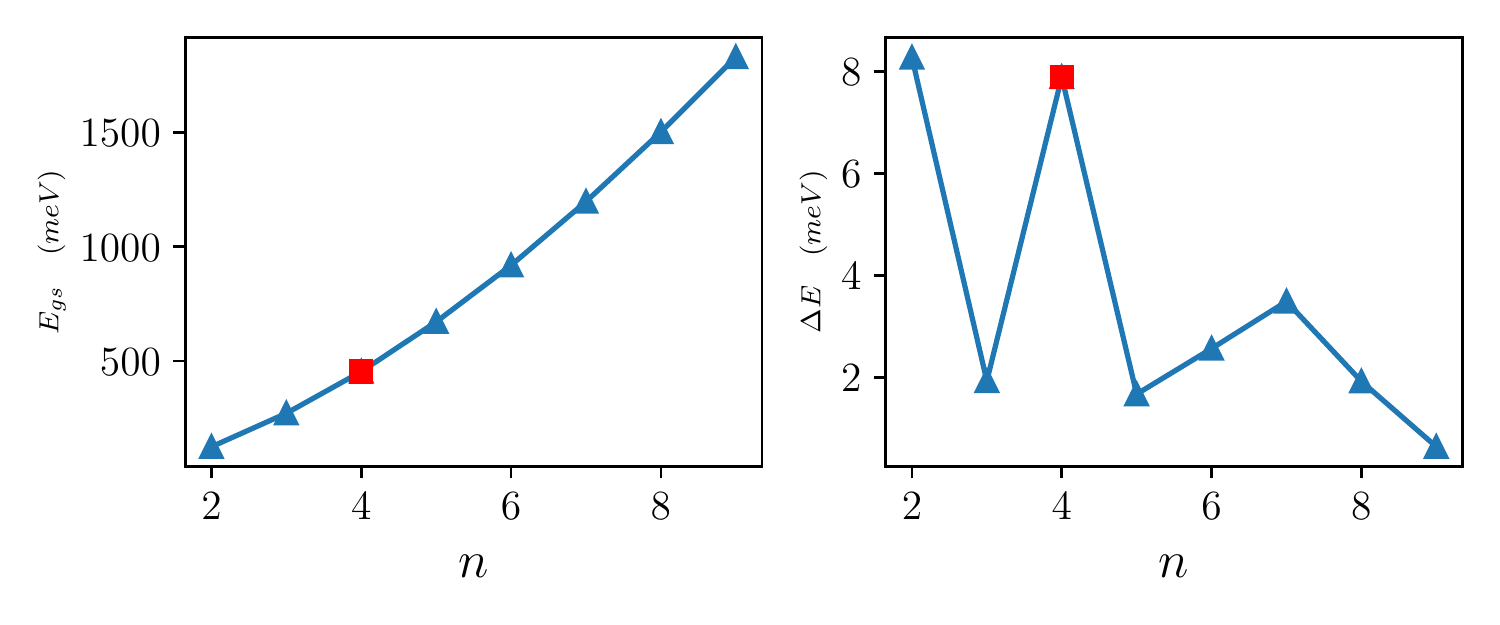}
    \caption{Ground state energy and and spin gap of multi-electron moir\'e atom  vs. the number of electrons $n$. The ground state has minimum total spin quantum number except $n=4$ (highlighted with red marker) which has $S=1$. ($\phi=30^{\circ}$, $\epsilon=10$, $V=20$meV, $a_M=15$nm, $m=1 m_e$)} 
\label{fig:moire610}
\end{figure}

\emph{2D Fermionic Neural Network ---}
To simulate the strongly correlated few-electron moir\'e atom, we utilize a 2D fermionic neural network based on FermiNet~\cite{ferminet}.
The many-body wave function is represented as
\begin{equation}
    \psi(\br) = \psi(r_1^{\uparrow},\dots,r_n^{\da}) = \sum_k \prod_{\sigma=\ua,\da} \textup{det}[\phi_i^{k\sigma}(r_j^{\sigma}; \{r_{/j}^{\sigma},r^{\bar{\sigma}}\})] \\
\end{equation}
where $k$ is the number of determinants,  
$\bar{\sigma}$ is the opposite spin of $\sigma$, $r_j^{\sigma}$ is the coordinate of the $j$-th electron with spin $\sigma$, $r_{/j}^{\sigma}$ is the set of coordinates of spin $\sigma$ electrons except for the $j$-th electron, and $r^{\sigma}$ is the set of coordinates of all spin $\sigma$ electrons.

The key of the FermiNet is to promote the standard single particle orbitals into many-particle orbitals $\phi_i^{k\sigma}(r_j^{\sigma}; \{r_{/j}^{\sigma},r^{\bar{\sigma}}\})$, which generalizes the neural network backflow transformation~\cite{Luo_2019} to continuous space (the details can be found in Ref.~\cite{ferminet}). In this work, we build a 2D fermionic neural network based on the FermiNet architectures with a few differentiating features. First, we work with 2D quantum materials so that all the coordinates have only two components instead of three compared to the conventional FermiNet. Second, our continuum model Hamiltonian does not involve nuclei so that the many-particle orbital functions purely depend on the electron coordinates. Third, we use an isotropic Gaussian envelope function in the many-particle orbital functions (see Supplementary Materials for details).

To optimize the fermionic neural network, we adopt the variational Monte Carlo (VMC) approach by minimizing the ground state energy.

\begin{equation}
    E(\theta) = \frac{\int d\br \psit^{*}(\br) H \psit(\br)}{\int d\br \psit^{*}(\br) \psit(\br)}
    \label{eq:energy}
\end{equation}
The optimization is performed using gradient descent with KFAC~\cite{martens2015optimizing}. The details can be found in the Supplementary Materials.

We have confirmed the accuracy of our method by comparison with previous literature on 
quantum dots with parabolic confinement. In Table.~\ref{tab:qdot}, we compare the ground state energies of $N$-electrons in a  parabolic quantum dot obtained from our 2D fermionic neural network calculation with the results of a configuration interaction calculation reported in Ref.~\cite{rontani2006full}. In all cases, FermiNet achieves a lower ground state energy (see Supplementary Materials).

\begin{figure}[t]
    \centering
\includegraphics[width=1\columnwidth]{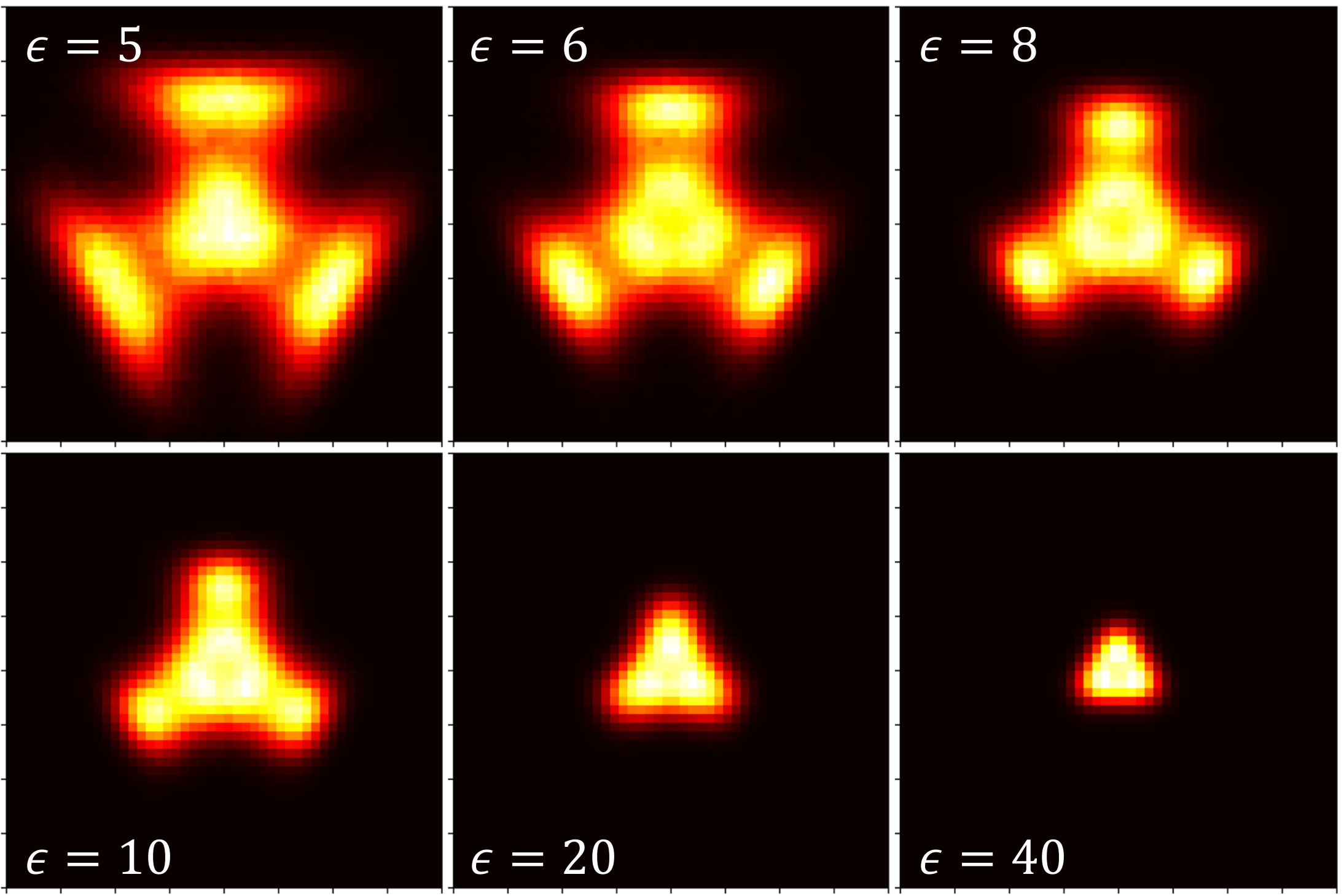}
    \caption{Ground state charge density of 6-electron moiré atom at several interaction strengths, showing the evolution from Wigner molecule at small $\epsilon$ to Schr\"{o}dinger atom at large $\epsilon$.  ($\phi=30^{\circ}$, $V=20$meV, $a_M=15$nm, $m=1 m_e$) 
    } 
\label{fig:WignerCrossover}
\end{figure}

\begin{figure*}[t]
\includegraphics[width=1.8\columnwidth]{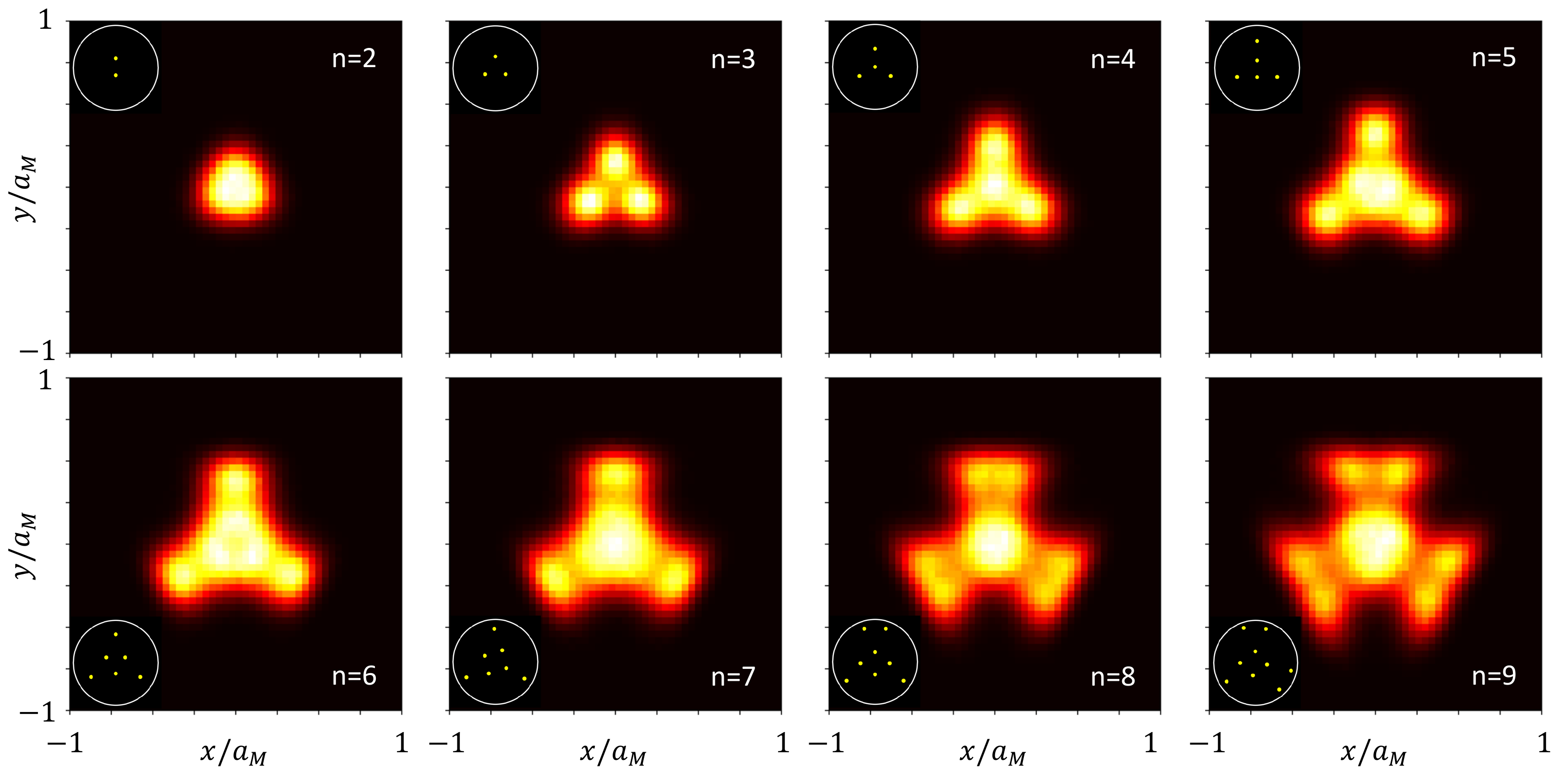}
\caption{Charge densities of $n$-electron moiré atom ground states for $n=2-9$ calculated with the 2D fermionic neural network and the corresponding classical ground states (insets). ($\phi=30^{\circ}$, $\epsilon=10$, $V=20$meV, $a_M=15$nm, $m=1 m_e$)}\label{fig:Fig1}
\end{figure*}

\emph{Results ---}
We perform simulations on \mo atoms using the 2D fermionic neural network for $n$ from 2 to 9 with different choices of $\epsilon$ and $\phi$. Fig. \ref{fig:moire610} shows the ground state energy as a function of electron number for $\phi=30^{\circ}$ and $\epsilon=10$. For all electron numbers except $n=4$, the ground state has minimum total spin $S=0$ for even $n$ and $S=1/2$ for odd $n$, whereas $n=4$ has $S=1$, consistent with Hund's rule that applies to the degenerate $p_x/p_y$ doublet. We note that the $n=4$ ground state of a parabolic quantum dot is also experimentally found to have $S=1$ \cite{tarucha1996shell}.

We also show the spin gap, defined as the difference in energy between the ground state and the first excited state with a larger total spin: $\Delta E_s \equiv E_{min}(S_{gs}+1) - E_{gs}$, where $S_{gs}$ is the total spin of the ground state. This quantity determines the magnetic field strength required to increase the moiré atom's total spin (assuming that orbital effect is small compared to the Zeeman effect).

Next, we investigate the charge density profile as a function of interaction strength 
and show the crossover from Schr\"{o}dinger atom to a Wigner molecule. 
Fig \ref{fig:WignerCrossover} shows the ground state charge density for the $n=6$ electron moir\'e atom for a range of dielectric constants $\epsilon$.  
When $\epsilon$ is large, the charge density is strongly concentrated at the origin as expected from the non-interacting limit.
For stronger, and more realistic, interaction strength $\epsilon\lesssim 10$, the charge density develops  distinct peaks away from the origin.  
These charge density peaks indicate the crossover to the Wigner molecule regime, and implies that electrons are ordered in a sequence of radial ``shells''.  
The $n=6$ Wigner molecule self-organizes into an inner and outer shell of 3 electrons each.  
The moir\'e crystal field clearly has a strong influence on the location of the charge density peaks, which form a distinct triangular shape.

To further investigate the shell structure in the Wigner molecule regime, we show the ground state charge density at $\epsilon=10$ 
and $\phi=30^{\circ}$ for a range of electron numbers $n=2-9$ in Fig~\ref{fig:Fig1}. 
 In the insets, we show the corresponding ground state of classical point charges (determined by minimizing the moiré potential and Coulomb energies). The first clear example of the Wigner molecule occurs at $n=3$, where the threefold symmetric crystal field pins an equilateral triangle of charge \cite{reddy2023moir}.
 In some cases, the classical ground states are unique ($N=3,4,6$) while, in others, there are several degenerate classical ground states related by point group operations. 
 In the classical limit $m\rightarrow\infty$ (equivalently $\hbar\rightarrow 0$), the quantum ground state is approximately equivalent to a superposition of classically degenerate ground states. The number of nearly degenerate classical states generally increases with $n$. At a finite, large mass, quantum fluctuations within the manifold of low-energy classical states result in a more homogeneous charge density profile.

\begin{figure}[t]
    \centering
\includegraphics[width=\columnwidth]{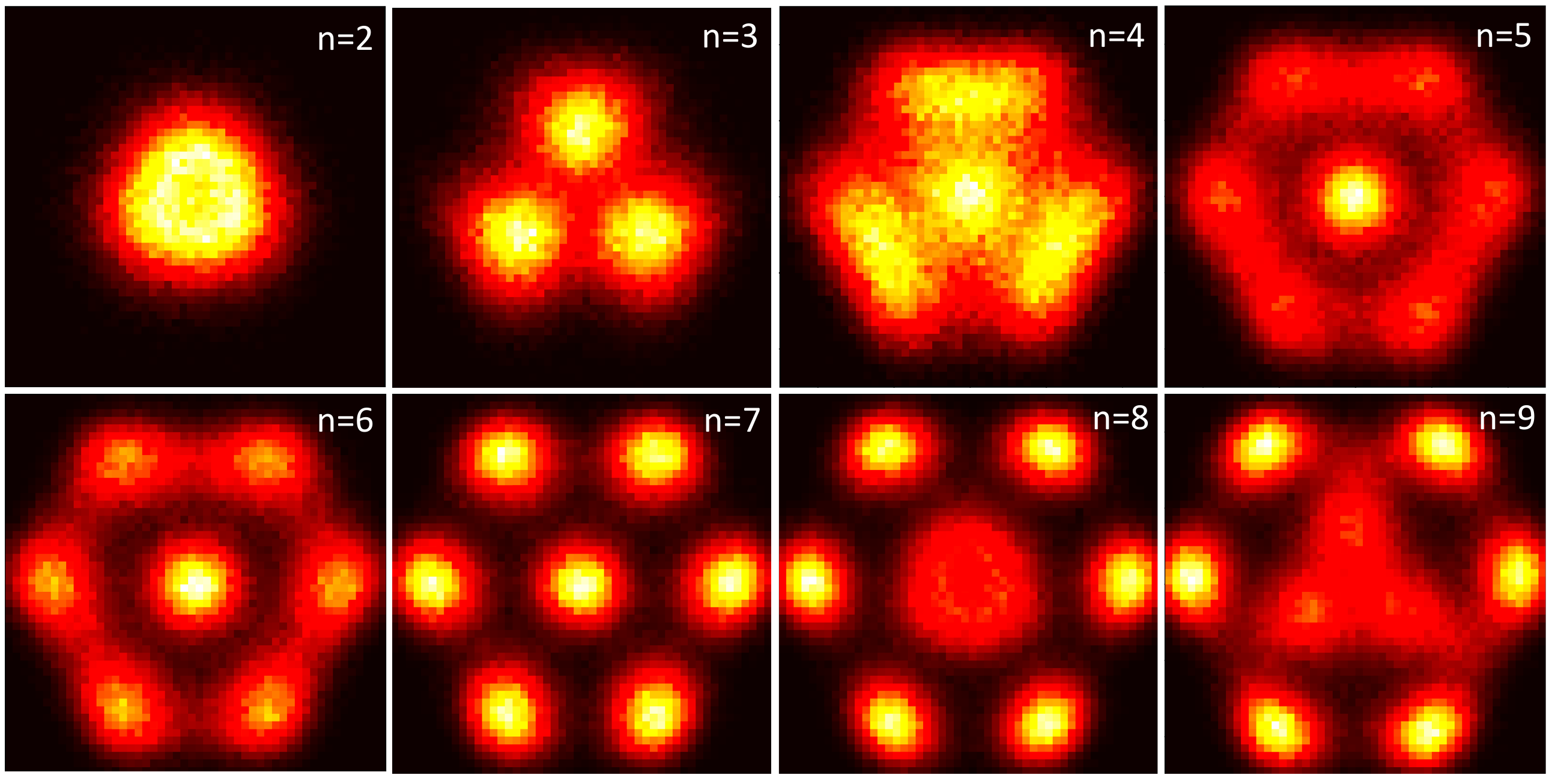}
    \caption{Charge density distributions of $n=2-9$ moiré atoms ($\phi=10^{\circ}$, $\epsilon\approx 3$, $V=15$meV, $a_M=14$nm, $m=0.5 m_e$)} 
\label{fig:phi10StrongInt}
\end{figure}

Comparing the results presented in Figs. \ref{fig:Fig1} and \ref{fig:phi10StrongInt}, we see that the charge density distribution depends strongly on the continuum model phase parameter $\phi$ through the moiré crystal field. In the two cases ($\phi=30^{\circ}$ vs $\phi=10^{\circ}$), the charge density distribution is similar for $n=2,3$ but differs significantly for higher $n$. For $\phi=10^{\circ}$, at $n=7$ and beyond, 6 electrons form an outer hexagonal shell and the remaining $n-6$ electrons form an inner shell that resembles the configuration of $n-6$ electrons alone.

\emph{Discussion--} Employing a novel 2D fermionic neural network approach, we have demonstrated that few electron moiré atoms exhibit ``Wigner crystal" charge density profiles driven by the interplay between Coulomb repulsion and the moiré crystal field. Our results pertain to semiconductor moiré superlattices at integer filling factors, where the isolated moiré atom model provides a good description of the system's ground state. Scanning tunneling microscopy is an ideal method for observing our predictions.

Our theory pertains most directly to heterobilayers with a large valence band offset such as WSe$_2$/WS$_2$ or $\Gamma$-valley homobilayers with strong inter-layer tunneling such as twisted bilayer WS$_2$, MoS$_2$ , and MoSe$_2$, which map to continuum models of the form Eq. \ref{eq:contham} \cite{wu2018hubbard, angeli2021gamma}.
In the absence of $D_6$ symmetry about the moiré potential minima, the moiré potential contours will generically exhibit a triangular anisotropy. Semiconductor moiré conduction bands often have higher effective masses $\sim 1 m_e$ than those of valence bands $\sim 0.5 m_e$, and for this reason their moiré atoms exist more deeply in the strong coupling regime. Lastly, we note that, unlike heterobilayers, moiré homobilayers do not have an upper bound on their moiré period due to a microscopic lattice constant mismatch, which is conducive to realizing the hierarchy of length scales $\xi_0 \ll \xi_c \ll a_M$ -- that is, the strong coupling limit of isolated moiré atoms.
Our 2D fermionic neural network approach can be applied a broad range of correlated electron phenomena in moiré materials and beyond.

\emph{Acknowledgements}--- 
It is a pleasure to thank Feng Wang and Hongyuan Li for a related collaboration, and Kin Fai Mak, Jie Shan, Ben Feldman, Marin Soljačić, Mike Entwistle, Norm M. Tubman, Gabriel Pescia, Wan Tong Lou and Cunwei Fan for helpful discussions. 
This work is supported by the Air Force Office of Scientific Research (AFOSR) 
under award FA9550-22-1-0432. 
DL acknowledges support from the NSF AI Institute for Artificial Intelligence and Fundamental Interactions (IAIFI) and the U.S. Department of Energy, Office of Science, National Quantum Information Science Research Centers, Co-design Center for Quantum Advantage (C2QA) under contract number DE-SC0012704. LF is partly supported by the David and Lucile Packard Foundation. 

\newpage

\bibliography{ref,ref_ml}

\appendix

\clearpage

\onecolumngrid
\begin{center}
	\noindent\textbf{Supplementary Material}
	\bigskip
		
	\noindent\textbf{\large{}}
\end{center}

\twocolumngrid

\section{Moiré potential and isolated atom model}

If the moiré potential $V(\bm{r})$ in Eq. \ref{eq:contham} is sufficiently smooth, it can be approximated with a Fourier expansion truncated to the lowest harmonics of the reciprocal superlattice, leading to the explicit form
\begin{equation}
V(\bm{r})=-2V_0\sum_{i=1,3,5}\cos(\bm{g}_i\cdot\bm{r}+\phi).
\end{equation}
Here, $\bm{g}_i=(\mathcal{R}_{2\pi/3})^{i-1}(0,\frac{4\pi}{\sqrt{3}a_M})$ are three low the shortest moiré reciprocal lattice vectors.
For the moiré atom model, we truncate the Taylor series expansion to $\mathcal{O}((r/a_M)^6)$, obtaining (in polar coordinates)
\begin{align}\label{eq:atomTaylor2}
    \begin{split}
        &V(r,\theta)/V_0 = -6\cos(\phi)+8\pi^2\cos(\phi)(r/a_M)^2\\
        &+\frac{16\pi^3}{3\sqrt{3}}\sin(\phi)\sin(3\theta)(r/a_M)^3-\frac{8\pi^4}{3}\cos(\phi)(r/a_M)^4\\
        &-\frac{16\pi^5}{9\sqrt{3}}\sin(\phi)\sin(3\theta)(r/a_M)^5\\
        &+\frac{16\pi^6}{405}\cos(\phi)(10-\cos(6\theta))(r/a_M)^6
    \end{split}
\end{align}
We choose to truncate at this order because it is the lowest order that both includes moiré crystal field effects and is bounded from below.
\section{Architecture and Optimization Details of 2D Fermionic Neural Network}

The neural networks that output the many-particle orbitals $\phi_i^{k\sigma}(r_j^{\sigma}; \{r_{/j}^{\sigma},r^{\bar{\sigma}}\})$ consist of $L$ layers of one-electron stream output $h_i^{l\sigma}$ and two-electron stream output $h_{ij}^{l\sigma \zeta}$, where $l$ is the layer index, $i,j$ are the electron index, $\sigma,\zeta$ are the spin index. The construction of $h_i^{l\sigma}$ and $h_{ij}^{l\sigma \zeta}$ utilize the FermiNet structure. 
The final output gives rise to the many-particle orbital
\begin{equation}
    \phi_i^{k\sigma}(r_j^{\sigma}; \{r_{/j}^{\sigma},r^{\bar{\sigma}}\}) = (c_{i}^{k\sigma} h_j^{L\sigma} + b_i^{k\sigma})  \pi_i^{k\sigma} \text{exp} (-|p_i^{k\sigma} r^{\sigma}_j|)
    \label{eq:orbtial}
\end{equation}
where $c_i^{k\sigma}$, $b_i^{k\sigma}$ are weights and biases to be optimized. Compared to the conventional FermiNet, we remove the nuclei dependence in the Gaussian envelop function in Eq.~\ref{eq:orbtial}. In addition, We only consider isotropic envelop function with parameters $\pi_i^{k\sigma}$ and $p_i^{k\sigma}$ to be optimized.

The gradient for the energy expression in Eq.~\ref{eq:energy} can be computed by  

\begin{equation}
    \nabla_{\theta} E(\theta) = \mathbb{E}_{p(\br)}[E_L(\br) - \mathbb{E}_{p(\br)}E_L(\br)] \textup{log}|\psit(\br)|] 
\end{equation}
where $E_L(\br)$ is the local energy defined by $E_L(\br)=H\psit(\br)/\psit(\br)$, $p(\br)=|\psit(\br)|^2$.
The parameter update for $\theta$ is done through Stochastic Reconfiguration~\cite{sorella1998green},
\begin{equation}
    \theta_{t+1} = \theta_t - \eta F(\theta_t) \nabla_{\theta_t} E(\theta_t)
\end{equation}
where $\theta_t$ is the parameter at step $t$, $\eta$ is the learning rate, $F(\theta_t)$ is the Fisher information matrix defined by

\begin{equation}
    F_{ij}(\theta) = \mathbb{E}_{p(\br)} [\frac{\partial \text{log} p(\br)}{\partial \theta_i} \frac{\partial \text{log} p(\br)}{\partial \theta_j}]
\end{equation}

In practice, we use the KFAC optimizer to approximate the above Fisher information matrix~\cite{martens2015optimizing}. The learning rate $\eta(t)$ is based on the scheduler $\frac{\eta_0}{1+t}$ with initial value $\eta_0 = 0.05$. For the simulations in the paper, we choose $k=4$, $L=4$, $hs_{dim}=256$ for one-electron stream networks, $hd_{dim}=32$ for two-electron stream networks. The total optimization step is 1600 and each stochastic gradient step uses 512 samples. No pretraining is required.

\section{Additional Data}

\subsection{Benchmark on Isotropic Potential}

We benchmark the 2D fermionic neural network with the configuration interaction results~\cite{rontani2006full} on the Hamiltonian in the isotropic potential limit (i.e. 2D quantum dot). The results are summarized as follows 

\begin{table}[h!]
\begin{tabular}{ | m{2em} | m{1cm}| m{1.1cm} | m{1.1cm}| m{1.1cm}| m{1.1cm}| m{1.1cm} | m{1.1cm}| m{1.1cm}  |} 
  \hline
  $n_e$ & 2 & 3 & 4 & 5 & 6 & 7 & 8 \\ 
  \hline
  CI & 4.8502 & 11.043 & 19.035 & 28.94 & 40.45 & 53.712 & 68.441 \\ 
  \hline
  FNN & 4.8487 $\pm 5e^{-4}$ & 11.0395 $\pm 2e^{-4}$ & 19.0290 $\pm 4e^{-4}$ & 28.9372 $\pm 9e^{-4}$ & 40.436 $\pm 2e^{-3}$ & 53.695 $\pm 1e^{-3}$ & 68.409 $\pm 2e^{-3}$\\ 
  \hline
\end{tabular}
\caption{Ground state energy benchmark.}
\label{tab:qdot}
\end{table}

$n_e$ is the number of electrons, CI stands for configuration interaction, FNN stands for the 2D fermionic neural network that we use. It is clear that the fermionic neural network consistently achieves lower energy and has better performance than the configuration interaction calculations. 

\subsection{Comparison with truncated ED for $n_e=2$}

\begin{figure}[h!]
    \centering
\includegraphics[width=0.8\columnwidth]{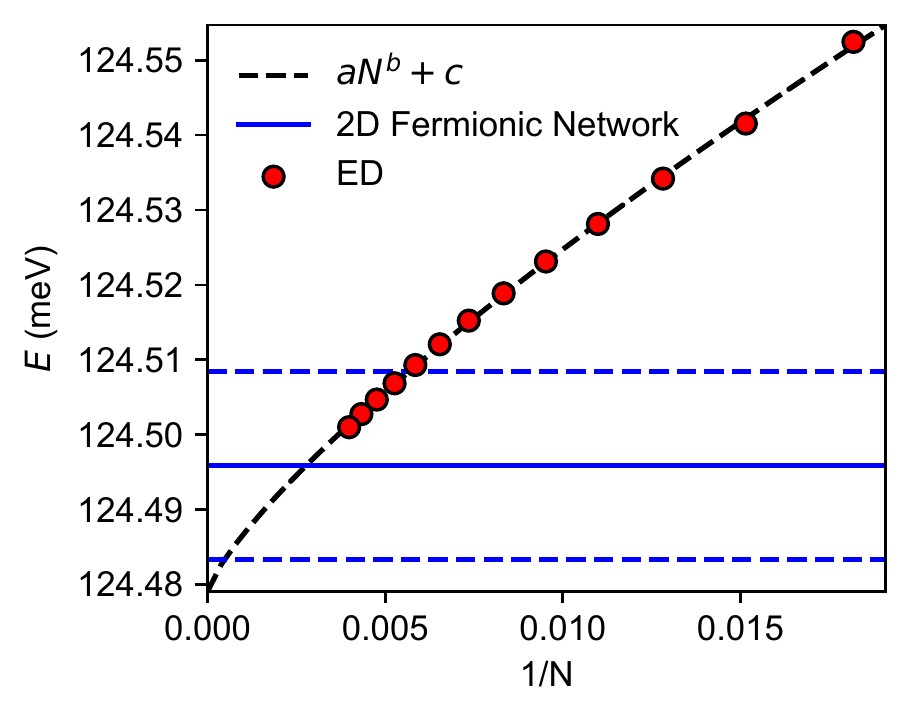}
    \caption{Comparison of the $n_e=2$ ground state energies obtained via truncated ED and 2D fermionic network calculations. Dashed blue lines indicate $\pm$ one standard deviation of the 2D fermionic network result. The black dashed line is a least squares fit to the ED results to the form $E(N) = aN^{b} + c$ with fit parameters $a=1.6591893166$meV, $b=-0.7802927277$, and $c=124.4791074426$meV. ($\phi=30^{\circ}$, $V=20$meV, $\epsilon=10$, $a_M=15$nm, $m=1 m_e$)}
    \label{fig:EDComparison}
\end{figure}

In Fig. \ref{fig:EDComparison}, we compare the results for the $n_e=2$ ground state energy obtained via an exact diagonalization calculation with those of our 2D fermionic network calculation. We plot the ground state energy obtained according to exact diagonalization with a truncated Hilbert space as a function of $N$, the number of single-particle states included in the truncated basis. The ground state energy obtained via fermionic neural network is lower than that obtained by exact diagonalization for all basis sizes we have checked. The extrapolated energy in the full basis limit falls nearly within one standard deviation of the 2D fermionic network result. The error bar of the 2D fermionic neural network calculation is based on 400 batches with each batch size of 512 samples.

\subsection{Charge Density of Different Model Parameters}

We provide additional charge density plots for different values of $\phi$ and $\epsilon$.

\begin{figure}[H]
    \centering
\includegraphics[width=\columnwidth]{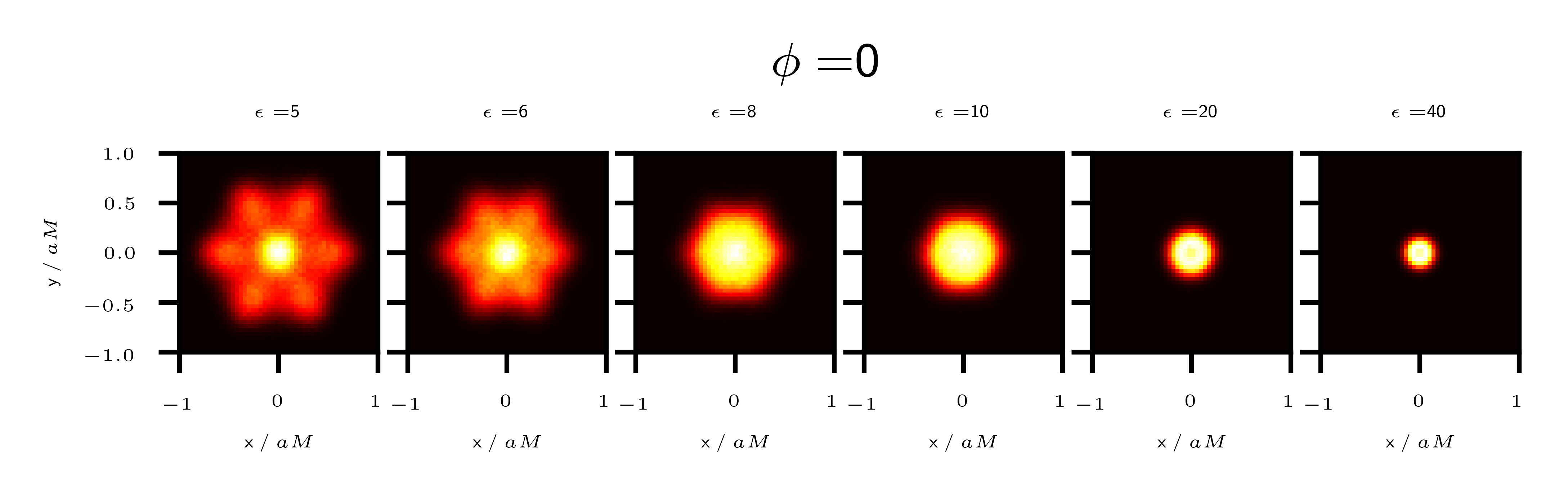}
    \caption{Moir\'e atoms , $\phi=0^{\circ}$ over $\epsilon=5,6,8,10,20,40$.} 
\label{fig:lambda}
\end{figure}

\begin{figure}[H]
    \centering
\includegraphics[width=\columnwidth]{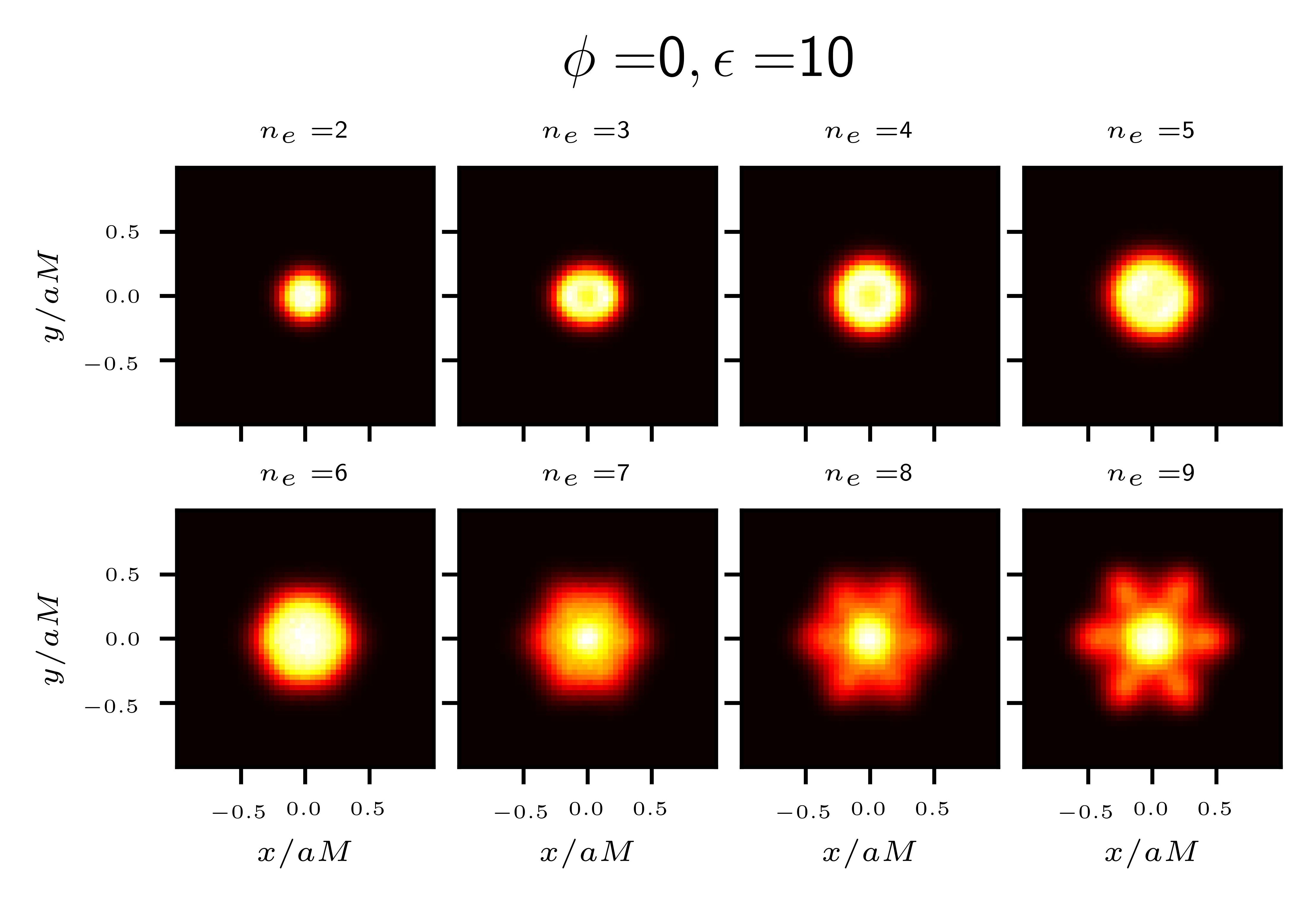}
    \caption{Moir\'e atoms density, $\phi=0^{\circ},\epsilon=10$. The rotation symmetry breaking of the $n_e=3$ state is a consequence of ground state degeneracy.} 
\end{figure}

\begin{figure}[H]
    \centering
\includegraphics[width=\columnwidth]{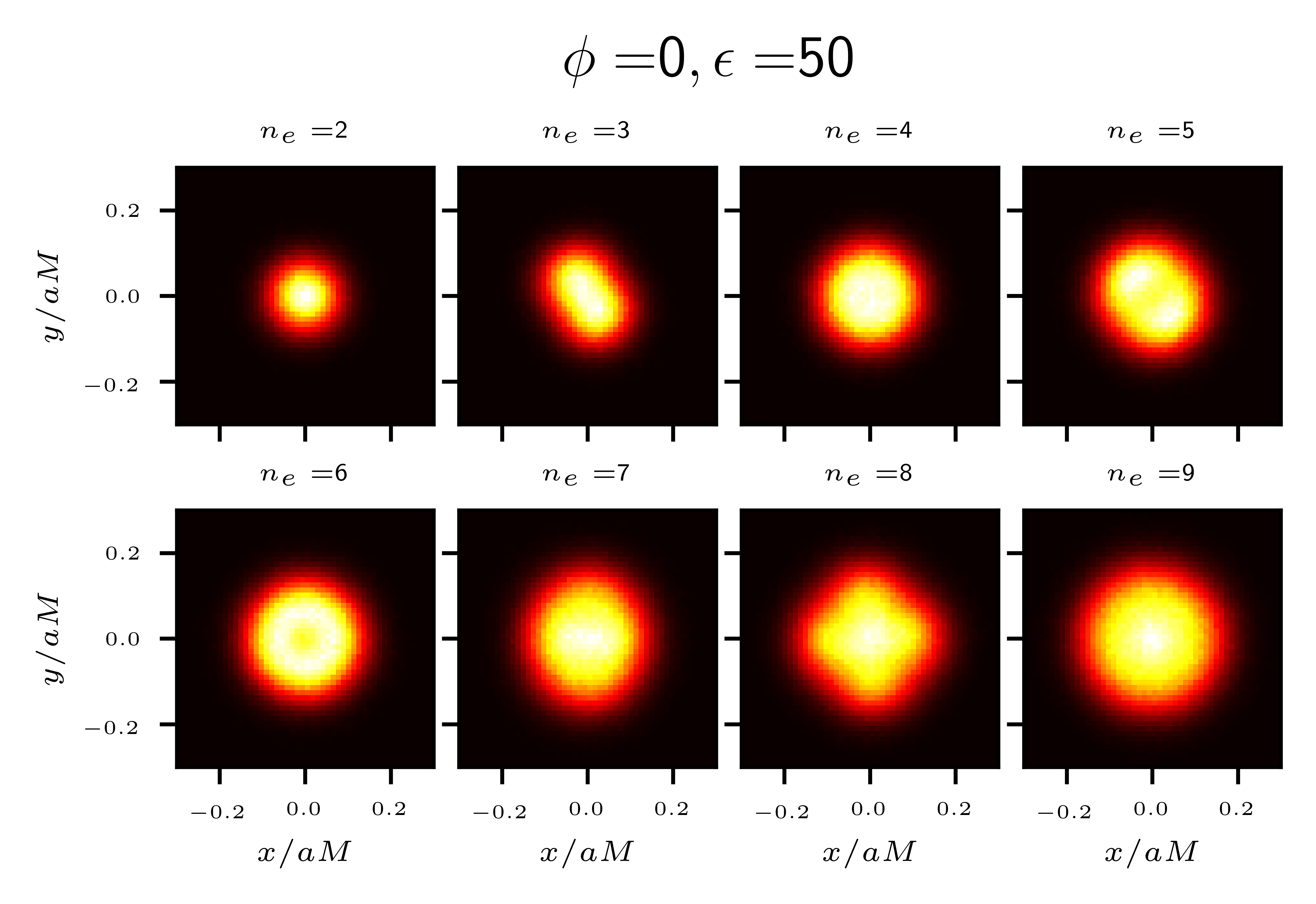}
    \caption{Moir\'e atoms density, $\phi=0^{\circ}$, $\epsilon=50$. The rotational asymmetry in the $n_e=3$ charge density is due to the ground state degeneracy, which can be understood from the electron configuration $1s^22p$ in the non-interacting limit.} 
\label{fig:moire100}
\end{figure}

\end{document}